\documentclass[11pt]{article}
\usepackage{times}
\usepackage{geometry}
\usepackage[utf8]{inputenc}
\usepackage{enumitem,amssymb}
\usepackage{graphicx}
\usepackage{fancyhdr}
\usepackage{aas_macros}
\usepackage{mdframed} 

\usepackage[authoryear]{natbib}
\setcitestyle{authoryear,open={(},close={)}}
\mdfdefinestyle{theoremstyle}{
innertopmargin=\topskip,}
\mdtheorem[style=theoremstyle]{lrptextbox}{}

\providecommand{\sorthelp}[1]{}
\title{Cosmic Magnetism}

\author{Jennifer West (University of Toronto) \and
Jo-Anne Brown (University of Calgary) \and
Bryan Gaensler (University of Toronto) \and
Alex S Hill (University of British Columbia, Dominion Radio Astrophysical Observatory) \and
Judith Irwin (Queen's University) \and
Roland Kothes (Dominion Radio Astrophysical Observatory) \and
Tom Landecker (Dominion Radio Astrophysical Observatory) \and
Tim Robishaw (Dominion Radio Astrophysical Observatory) \and
Samar Safi-Harb (University of Manitoba) \and
Jeroen Stil (University of Calgary) \and
Cameron Van Eck (University of Toronto) \and
Gregg Wade (Royal Military College of Canada)}

\begin{document}
% ****BEGIN MAIN WHITE PAPER SECTION****
\setcounter{page}{0}

\maketitle

\section*{Executive summary}
%\section*{EOI}
Magnetic fields are involved in every astrophysical process on every scale: from planetary and stellar
interiors to neutron stars, stellar wind bubbles and supernova remnants; from the interstellar medium in
galactic disks, nuclei, spiral arms and halos to the intracluster and intergalactic media. They are
involved in essentially every particle acceleration process and are thus fundamental to non-thermal
physics in the Universe. Key questions include the origin of magnetic fields, their evolution over
cosmic time, the amplification and decay processes that modify their strength, and their impact on
other processes such as star formation and galaxy evolution. Astrophysical plasmas provide a unique
laboratory for testing magnetic dynamo theory. The study of magnetic fields requires observations that
span the wavelength range from radio through infrared, optical, UV, X-ray, and gamma-ray.

Canada has an extremely strong record of research in cosmic magnetism, and has a significant
leadership role in several ongoing and upcoming global programs. This white paper will review the
science questions to be addressed in the study of cosmic magnetic fields and will describe the
observational and theoretical opportunities and challenges afforded by the telescopes and modelling
capabilities of today and tomorrow.

\newpage

\section{Introduction}

Magnetic fields are ubiquitous in space, playing what must be crucial, but often poorly understood, roles in many astrophysical processes. Magnetic fields span many orders of magnitude in both their physical scale and field strength reaching as high as $10^{15}$~Gauss in magnetars to as low as $10^{-9}$~Gauss in intergalactic regions. Since they do not radiate and cannot be observed directly, their study is challenging. Canadians have a long history leading studies to answer many important questions about magnetism in the cosmos. In the last decade, more than 30\% of all refereed astronomy papers, with Canadian contributions, refer to ``magnetism'' or ``magnetic fields'' (ADS).

Canadian contributions include the first detections of magnetic fields in white dwarfs \citep{1970ApJ...161L..77K}, pre-main sequence Herbig Ae/Be stars \citep{2005A&A...442L..31W}, and evolved post-AGB stars \citep{2015MNRAS.446.1988S}, the first identification of the magnetic field reversal between the local arm and the Sagittarius arm \citep{1979Natur.279..115S}, the first detection of magnetic fields in high-redshift objects \citep{1982ApJ...263..518K}, the first detection of a magnetic field within a cluster of galaxies \citep{1986A&A...156..386V}, the largest ever catalogue of extragalactic rotation measures \citep{2009ApJ...702.1230T}, the best map of Galactic Faraday rotation \citep{2012A&A...542A..93O}, the best model of the large-scale structure of the magnetic field in the disk of the Milky Way \citep{2011ApJ...728...97V}, the best detailed maps of polarized dust by Planck \citep[e.g.,][]{collaboration2018planck}, the first broadband all-sky survey of radio polarization \citep{2019AJ....158...44W}, and fundamental new processing algorithms such as the polarisation gradient \citep{2011Natur.478..214G,2017MNRAS.466.2272H}, polarisation stacking \citep{2015ASPC..495..241S} and real-time ionospheric Faraday correction. Canadians are also world leaders in the development of techniques and technology related to radio magnetism studies, such as telescope dishes, receivers and correlators. 

\section{\label{sec:obs}Observational techniques}
Radiation of relativistic particles in the presence of magnetic fields produces intrinsically polarized synchrotron radiation making radio observations, and in particular radio polarization, especially useful for probing magnetic fields. Polarized starlight optical wavelengths and dust grains at mm wavelengths provide useful magnetic field tracers in other wavebands. High-energy observations provide additional, complementary information about cosmic ray populations that can inform magnetic field studies. Some of these topics are covered in other white papers including: E025 (star formation), E076 (dust), and E081 (interstellar medium, ISM).

Synchrotron emission and Faraday rotation occur along virtually all sight lines through the Galaxy. The parameter that characterizes the medium to a distance $d$ is the Faraday depth 
%\begin{equation}
  $\phi(d) = {0.812 \,} \int_d^{\rm{telescope}} {n_e(r)} {B_{||}(r)} {\rm{d}l}$
 %\end{equation}
where $n_e$[cm$^{-3}$] is the electron density, $B_{||}$[$\mu$G] is the
line-of-sight component of the magnetic field, and ${\rm{d}}l$[pc] is the distance along the line of sight.

In the simple case of a polarized extragalactic source seen through the Galactic disk, or a pulsar seen through part of the disk, the Faraday depth becomes the Rotation Measure (RM). For this case
polarization angle, ${\theta}$, is a linear function of ${{\lambda}^2}$ and RM is relatively simple to measure; extensive surveys of point-source RMs have been used to great effect to map the large-scale structure of the Galactic magnetic field. When emission and rotation are mixed, $\phi$ is no longer proportional to ${\lambda}^2$ and interpretation of polarization data on the extended emission becomes more complicated. Different Faraday depths can occur at different distances along the line of sight, and the true situation is portrayed by the Faraday depth spectrum, produced by applying Rotation Measure Synthesis. The resolution in Faraday depth depends mostly on the longest wavelength of the data. The maximum width of the Faraday depth structures that can be successfully mapped depends mostly on the shortest wavelength. In other words, combining low frequencies with wide bandwidths is the key to successful investigation of the magneto-ionic medium. 
This is key when choosing an observing band or instrument for Faraday rotation studies.  

\section{\label{sec:science}Key Science Questions}

With significant involvement and leadership in many current and upcoming international magnetism related projects and telescopes, Canadians are in a position to make significant contributions to many fundamental questions in magnetism science. We present a summary of some of the questions for which significant Canadian contributions for advancement can be expected over the next decade.

\textbf{How do magnetic fields influence stellar evolution?} 
Magnetic fields are a natural consequence of the dynamic plasmas that comprise a star. They directly and indirectly impact stellar lives through modification of convective and circulatory interior flows, redistribution of angular momentum and nucleosynthetic chemicals, channeling and modification of mass loss and accretion, and shedding of rotational angular momentum through magnetic braking. Ultimately, these effects lead to important modification of stellar evolutionary pathways \citep[e.g.,][]{2018CoSka..48..124K} and stellar feedback effects \citep[e.g.,][]{2005ApJ...626..350H}, such as mechanical energy deposition in the ISM and supernova explosions, and hence the properties of stellar remnants and the structure and chemistry of the local Galactic environment.

\textbf{What is the magnetic field in ISM drivers such as supernova remnants and molecular clouds?} Magnetic fields pervade in the interstellar medium and are believed to shape the process of star formation, yet probing magnetic fields is challenging. Zeeman splitting can be used to measure the total magnetic field and Faraday rotation measurements of background sources can be used to find the direction and magnitude of the component of magnetic field along the line-of-sight to star forming regions \citep{2018A&A...614A.100T}.

The blast wave from supernova explosions expands to large scales, sweeping up and compressing the ambient magnetic field making supernova remnants (SNR) excellent probes for local structures in the Galactic mean field \citep{1998ApJ...493..781G,2016A&A...587A.148W}. However, broadband fits, X-ray observations and 3D simulations of SNR including efficient particle acceleration show evidence for additional magnetic field amplification at SN shocks
\citep[e.g.,][]{2007Natur.449..576U,  2014ApJ...789...49F}. The exact 3D structure and strength of SNR magnetic fields, particularly in the early phases of their evolution, remains unclear. 

\textbf{What is the small scale structure of the Galactic magnetic field?} Turbulent magnetic fields are thought to be a significant component of the Galaxy with a magnitude equal to or greater than the mean field component \citep{2015ASSL..407..483H}. Recent studies have shown correlations between neutral hydrogen filaments and the magnetic field alignment \citep{2018ApJ...857L..10C} as observed with diffuse dust emission and starlight polarization. There is much still unknown about the turbulent properties of the Galactic magnetic field including the scales, ratio of random to regular components, the nature of turbulence (i.e., isotropic vs anisotropic random components), and whether the field has helicity \citep{2011A&A...530A..89O}. 

\textbf{What is the large scale 3D magnetic field structure of the Milky Way Galaxy?} Observations of nearby spiral galaxies have revealed a regular large scale pattern that follows the spiral arms. However these observations are 2D projections of a 3D field, of which the exact topology remains a mystery \citep{Collaboration:2016eh}. Understanding the origins and evolution of galactic magnetic fields in general require this understanding. Our Milky Way provides us with a unique perspective to probe the large scale field from the inside. Studies to date have revealed probable field reversals in the arms but provide an incomplete picture \citep{2007ApJ...663..258B,2011ApJ...728...97V,2017A&A...603A..15O}. With new and better data, the next decade should see significant advancement in the development of a trustworthy model of the Galactic magnetic field \citep{2018JCAP...08..049B}, which will in turn allow us to properly subtract it to reveal the extra-galactic sky in more detail.

\textbf{What is the 3D magnetic field structure of nearby Galaxies?} 
An important outstanding scientific issue regarding galactic halos and their magnetic fields relates to {\it lagging halos}, i.e. the fact that the rotation of the halo lags behind the rotation of the disk.  Just why this occurs and how the lag may 'connect' to the IGM are not yet known although magnetic fields could be the missing link \citep[e.g.][]{2016MNRAS.458.4210H}.  Another is, 'where do magnetic fields close?'  One would expect that the field lines close at some point, but current observations are not able to detect magnetic fields beyond a few kpc from the disk.  Others are: to what extent are magnetic fields affecting local or global dynamics? and ultimately, how are the fields generated?

Galactic magnetic fields should not be treated as minor perturbations, but rather as key ingredients in a rich dynamically active environment.  Probing deeper into their structure and physical state will provide answers to some of the key questions facing galaxy formation and evolution today.  Future instruments should improve on sensitivity by at least a factor of 10, while ensuring that a variety of spatial scales can be detected.

\textbf{What role do magnetic fields play in cosmic ray acceleration and propagation?} Galactic magnetic fields are thought to be responsible for the  acceleration of electrons, positrons and ions to cosmic rays. Within the Milky Way, this acceleration is thought to occur primarily in supernova remnants \citep[e.g.,][]{2017arXiv170608275M}. Cosmic rays propagate through the Galaxy mostly along field lines, but also by diffusion and advection \citep{2019arXiv190703789S}. Additionally, ultra-high energy cosmic rays (UHECRs) are extragalactic cosmic rays with energies exceeding $10^{18}$~eV. Acceleration to these extreme energies could be due to transients such as massive supernovae or compact object mergers, active-galactic nuclei, or galaxy clusters, though the exact mechanism is unclear. Understanding the creation and propagation of cosmic rays will be advanced through accurate knowledge and diagnosis of the magnetic fields in supernova remnants, and also through understanding of Galactic, extra-galactic, and intergalactic magnetic field strength and structure.

\textbf{How do magnetic fields affect AGN feedback?} Active Galactic Nuclei, powered by accretion on a supermassive black hole, eject relativistic jets that interact with the interstellar and intergalactic medium on sub-pc to Mpc scales. Magnetic fields in AGN are observed from pc scales to Mpc scales, e.g. from the non-thermal filaments in the Galactic centre, VLBI polarimetry of radio galaxy cores, to polarization of jets, and radio lobes. They affect the accretion disk, jet collimation and particle acceleration, and contribute significantly to the pressure inside radio lobes. Expanding radio lobes inject large amounts of energy in the intergalactic medium \citep[e.g.,][]{2007ARA&A..45..117M} that affects its dynamics and indirectly accretion of gas on galaxies. Polarized radio emission and Faraday rotation reveal the magnetic field structure in AGN and their interaction with the intergalactic gas.

\textbf{How have magnetic fields evolved over cosmic time?} Much remains unknown about the evolution of magnetic fields and the state of fields in the early Universe. Did the fields grow steadily over time? Or is there a phase of rapid field amplification? Recent studies show little to no change in the rotation measures between high and low redshift galaxies \citep{2007MNRAS.375.1059B,2018MNRAS.475.1736V}. However, sample sizes of high redshift polarized sources are low. Currently there are only approximately 20 radio galaxies with polarization properties and redshifts $z \ge 3.5$ that are known. New surveys such as the Polarisation Sky Survey of the Universe’s Magnetism (POSSUM) and the Very Large Array Sky Survey (VLASS) will probe deeper than ever to sample the largest number of high-redshift galaxies with radio polarization observations to date. This will in turn allow us to investigate the evolution of their Faraday rotation measures, which probes their magnetic fields and the electron densities of their local environments, over cosmic time. 

\textbf{What is the role of cosmic magnetic fields in large-scale structure formation and evolution?} We know that on the largest scales there is structure to the Universe: voids, filaments, and clusters (e.g. the cosmic web). Theory tells us the magnetic fields pervade all of this intergalactic space, but as yet there have been no measurements of intergalactic magnetic fields. The origin of cosmic fields and their evolution and role in structure formation is unknown. From simulations the strength of these fields can vary from nG to $\mu$G levels \citep[e.g.][]{2008Sci...320..909R,2014MNRAS.439.2662V,2016MNRAS.459...70V}, depending on the location of the field (e.g. voids or clusters) but also on the assumed strength of any primordial magnetic field and on the interactions and injections from galaxies and galaxy evolution. Recent statistical studies \citep{2017MNRAS.467.4914V,2017MNRAS.468.4246B} obtained upper limits on the field strength from new radio data, and the first direct detection of a intergalactic cosmic filament by the Low-Frequency Array (LOFAR) telescope was made earlier this year \citep{2019Sci...364..981G}. Upcoming surveys with existing and new instruments will provide unparalleled opportunities for advancement in this field.

\section{Current Canadian Leadership in the International Magnetism Community}

 Canadians currently have leadership roles in nearly all of the major ongoing international radio polarization surveys, addressing a wide cross-section of significant science questions. Canadian led projects include next generation Faraday rotation measure grid experiments  through POSSUM and VLASS, studies of the diffuse Galactic polarized emission in the Global Magneto-Ionic Medium Survey (GMIMS), and detailed studies of Galactic halo magnetic fields with CHANG-ES: Continuum Halos in Nearby Galaxies - an EVLA Survey. A recent \$10-million grant from Canadian Foundation for Innovation (CFI) and provincial partners for the Canadian Initiative for Radio Astronomy Data Analysis (CIRADA) will enable Canadians to develop the infrastructure and expertise needed to convert the enormous raw data streams from next-generation telescopes into enhanced data products that astronomers can directly use to make new discoveries. A significant component of this project is for cosmic magnetism related science with POSSUM and VLASS. In addition, Canadians also participate in projects on other instruments such as LOFAR, the Murchison Widefield Array (MWA), Australian Square Kilometer Array Pathfinder (ASKAP) and MeerKAT telescope, which probe deeper than ever before at a range of frequencies.
 
 At optical wavelengths, Canadian leadership in the Magnetism in Massive Stars (MiMeS), Binarity and Magnetic Interactions in Stars (BinaMIcS), and related projects continue to exploit the international suite of precision optical polarimeters on 4-metre to 8-metre class telescopes to drive forward our understanding of the evolutionary impact of magnetic fields in non-degenerate stars and white dwarfs.
 
High-energy studies provide insights into particle acceleration operating in SNRs, pulsar wind nebulae and active galactic nuclei and also help address the bigger questions of cosmic magnetism and the origin of high-energy cosmic rays, in synergy with studies at lower energies. These questions in turn are driving future telescopes in the radio (Square Kilometer Array, SKA), submillimetre (next generation JCMT camera), X-ray (ATHENA and new X-ray polarimeters such as eXTP and IXPE) and gamma-ray bands (Cherenkov Telescope Array).

\subsection{CIRADA: Canadian Initiative for Radio Astronomy Data Analysis}
Through CIRADA, Canadians will develop expertise necessary for management of large scale radio surveys. Current cutting-edge telescopes, like ASKAP, produce high resolution and multi-frequency data with volumes that are now at a point where it is often impossible for an individual astronomer to use a desktop computer to process and analyze these data on their own. Instead, supercomputers are required to make images and transform these into scientifically useful catalogues and advanced image products such as Faraday cubes. CIRADA is developing the pipeline for the creation of all of the advanced polarization catalogues and data products for VLASS and POSSUM. Together these will make a polarized radio map of the entire sky in unprecedented detail. 

\subsubsection{POSSUM: Polarisation Sky Survey of the Universe’s Magnetism}
POSSUM is one of ten major Survey Science Projects to be undertaken on ASKAP \citep{2010AAS...21547013G}. ASKAP is a radio telescope array located in Western Australia, which uses 36 antennas equipped with advanced receivers known as phased array feeds. It is an ideal survey instrument due to its wide field of view and fast mapping speed.  POSSUM will team up with other major science projects, the Evolutionary Map of the Universe (EMU) and Widefield ASKAP L-Band Legacy All-Sky Blind Survey (WALLABY), with commensal observations to maximize scientific output and to obtain wide frequency coverage in the range $\sim$800-1800~MHz. It will survey the entire sky (south of $\delta=+30^\circ$), to an RMS sensitivity of 10~$\mu$Jy/beam at 10$''$ resolution.

The main science result will be a catalogue of Faraday rotation measures (RMs) for around a million extragalactic radio sources at an unprecedented density of approximately 25-30 RMs per deg$^2$. We will also produce advanced products like catalogues of Faraday components, descriptions of Faraday complexity, and Faraday cubes to provide additional spatially resolved information. Such a dense RM-grid will allow us to probe magnetic features in the Galaxy, to better determine the 3D geometry of the Milky Way's magnetic field, to test dynamo theory and other models that describe the generation of large-scale magnetic fields, and to understand how magnetic fields have evolved as a function of redshift in galaxies, clusters and intergalactic medium (see~Sec.~\ref{sec:science} for more details).

\begin{figure*}[!ht]
\centering 

\begin{minipage}{8.5cm}
\includegraphics[width=7.3cm]{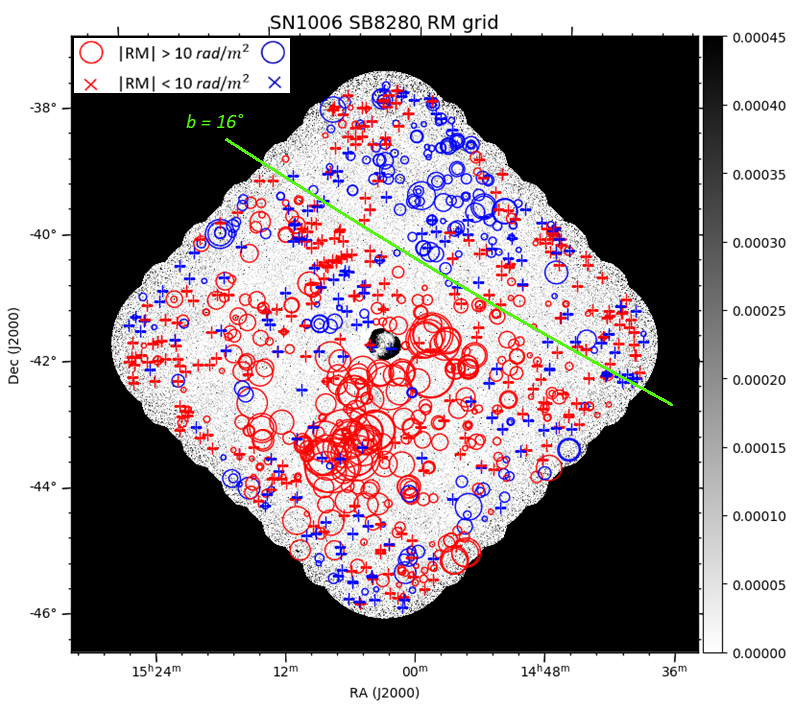}
%figures/m1-wT.pdf
\begin{scriptsize}\caption{ \label{fig:rmgrid}Preliminary RM-grid for a test POSSUM observation 
(Vanderwoude/POSSUM collaboration, in prep.) with 1040 polarized sources and a density of $\approx$~29 sources/deg$^2$. The previous best RM-grid in this region \citep{2019MNRAS.485.1293S} has only 12 sources.}
\end{scriptsize}
\end{minipage}
\hfill
\begin{minipage}{8.5cm}
\includegraphics[width=7.5cm]{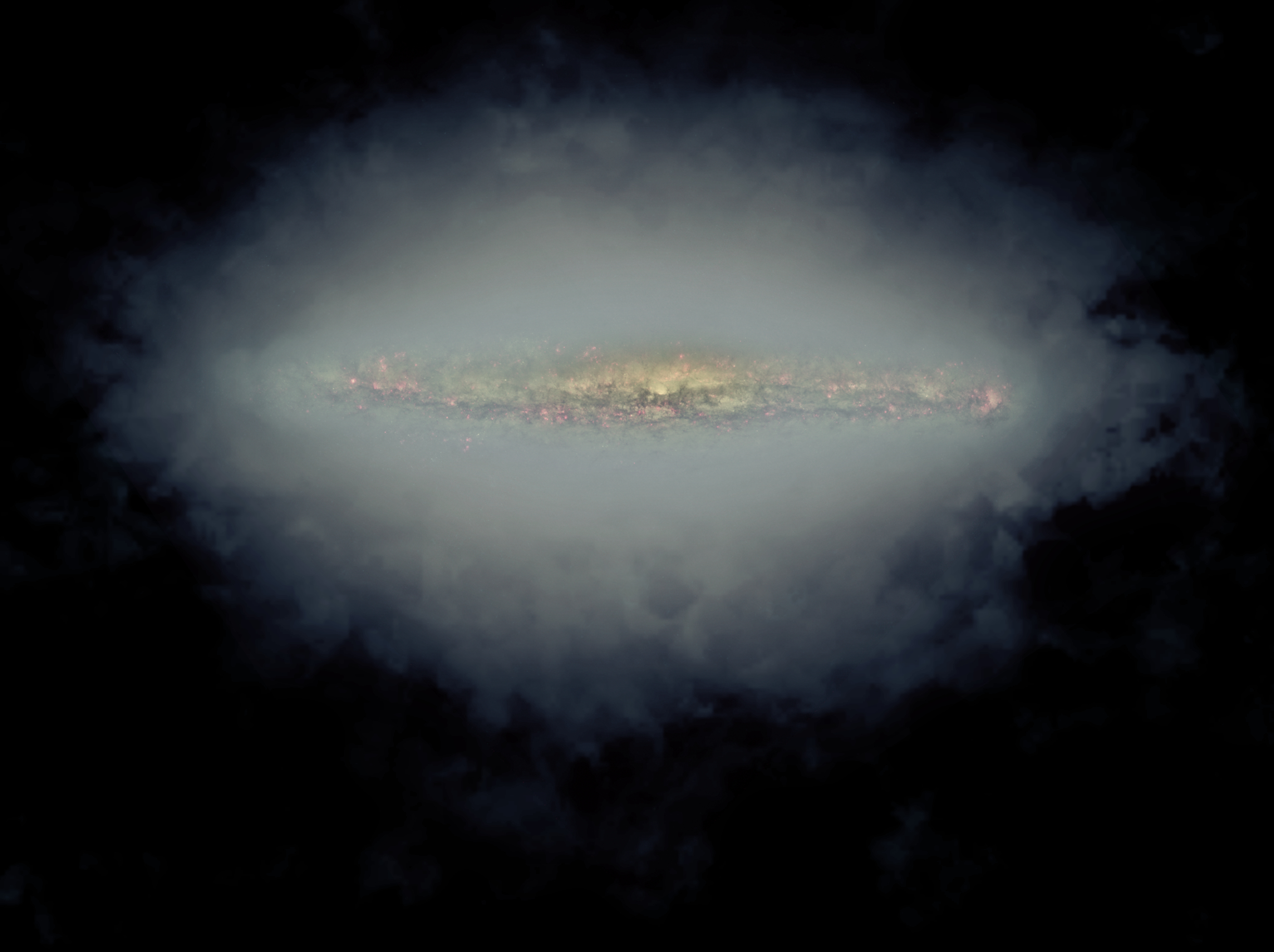}

\caption{ \label{fig:median}The median synchrotron halo of an edge-on spiral galaxy (in blue-grey) in L-band made from stacking 30 of the CHANG-ES galaxies superimposed on an optical Hubble Space Telescope image of NGC~5775. From \cite{2019AJ....158...21I}.
}

\end{minipage}

\end{figure*}

A number of test fields have been observed and are already showing very exciting and promising results, including the densest RM-grid ever produced (see Fig.~\ref{fig:rmgrid}), which allow us to tease out details of the Galactic magnetic field geometry on $\sim$pc scales. A full-scale pilot covering $\sim300$~deg$^2$ is currently underway with the full survey expected over the next five years. Ten Canadian astronomers are members of the POSSUM team, including four in leadership roles.

\subsubsection{VLASS: Very Large Array Sky Survey}
VLASS is a radio sky survey offering a unique combination of high angular resolution ($\approx2.5''$), sensitivity (a $1\sigma$ goal of 70~$\mu$Jy/beam in the coadded data), full linear Stokes
polarimetry, time domain coverage, and wide bandwidth (2–4 GHz) \citep{2019arXiv190701981L}. Observations will take place over three epochs to allow the discovery of variable and transient radio sources, for a total of 5500 hours to be observed on the Karl G. Jansky Very Large Array (VLA). Observations began in September 2017 and will continue until 2024 with the first epoch of observing now complete. VLASS covers the whole sky visible to the VLA (declination $> -40^\circ$), a total of 33 885 deg$^2$. 

Faraday Tomography of The Magnetic Sky is one of four key science themes addressed by the survey. It is estimated that $200,000$ sources with Faraday rotation measures will be found (almost 6 times that of the current largest known catalogue of RMs). The sky coverage will overlap with a section of the POSSUM survey, which, with the different frequency coverage and improved spatial resolution of VLASS, will improve the ${{\lambda}^2}$ coverage and thus provide better RM measurements (see Sec.~\ref{sec:obs}).

\subsection{GMIMS: The Global Magneto-Ionic Medium Survey}

GMIMS is an international project with 30 members in 9 countries, 11 of whom are in Canada. The goal of 
GMIMS is to improve our understanding of the Galactic magnetic field by mapping polarized radio emission over the entire sky, in the Northern and Southern hemispheres, using large single-antenna radio telescopes around the world.

 Although there are existing all-sky surveys, they exist only in narrow, widely spaced frequency bands. These data are inadequate for the characterization of Faraday depth, the main determinant of the appearance of the polarized radio sky at long wavelengths. GMIMS plans for complete coverage of the frequency range 300 to 1800 MHz with thousands of frequency channels. This is a crucial frequency range in terms of depolarization and Faraday depth coverage: at lower frequencies Faraday rotation so dominates that only quite local phenomena can be probed. At higher frequencies Faraday rotation is so weak that huge bandwidths are required. Rotation Measure Synthesis and other RM estimation techniques (e.g. QU fitting) are being used to analyze the data. GMIMS is the first project to apply Rotation Measure Synthesis to single-antenna data.
 
 For technical reasons the band has been divided into three segments, 300--800\,MHz, 800--1300\,MHz, and 1300--1800~MHz, the Low, Mid, and High bands.The sky naturally divides into North and South, so a total of six all-sky surveys are required to complete the dataset. Observations are complete for three surveys, one in the North and two in the South, and data reduction is mostly complete.

\setlist{nosep}
\begin{itemize}
    \item{ High band north: The DRAO Galt Telescope (26-m) 1270-1750 MHz, all RA, declination range $-30^\circ < \delta < +87^\circ$. Data reduction is virtually complete and four science papers have been published }%(Fig.~\ref{fig:gmims}).}
    \item{ Low band south: Parkes 64-m Telescope, 300-480 and 660-870 MHz, all RA, $-90^\circ < \delta < +20^\circ$. The 300--480~MHz data are published \citep{2019AJ....158...44W} and  available at the CADC. The upper part of the band ($> 480$~MHz) was heavily affected by radio-frequency interference (RFI). Two science papers on this survey have been published.}
    \item{High band south: Parkes, 1300-1800 MHz, all RA, $-90^\circ < \delta < 0^\circ$.
Data reduction is 90\% complete.}
\end{itemize}

To date, five science papers have been published from GMIMS data \citep{2010ApJ...724L..48W, 2015ApJ...811...40S, 2017MNRAS.467.4631H, 2019ApJ...871..106D, 2019MNRAS.487.4751T}. The GMIMS all-sky Faraday cubes are without precedent. It is evident for the first time that there is significant emission at non-zero Faraday depths. The published GMIMS papers explore fundamentally new analysis approaches to these data. The papers demonstrate the richness of the Faraday sky, and have provided clues to the structure of the magneto-ionic medium and the magnetic field configuration within the Galaxy. However, the real promise of GMIMS will be realized by combining the full range of frequencies to enable simultaneous resolution of small ($\sim 1 \textrm{ rad m}^{-2}$) features and sensitivity to large ($\sim 100 \textrm{ rad m}^{-2}$) features. Additional surveys are planned to achieve this goal.
An all-sky survey for Low-band North will be made with the 15-m DVA telescope at DRAO in 2020. This will subsequently be combined with CHIME data from 400--800\,MHz, and even later with CHORD data, to achieve sub-degree angular resolution. A proposal for Mid-band South data using the Parkes Telescope is being prepared, in collaboration with ASKAP projects POSSUM and EMU. A new survey with the DRAO Galt Telescope covering 900--1700 MHz is planned.

\subsection{CHANG-ES: Continuum Halos in Nearby Galaxies - an EVLA Survey}
CHANG-ES has 45 members in 8 countries, 9 of whom are in Canada. With over 400 hours of VLA time in 3 different array configurations (plus 200 hours of GBT time), and all 4 Stokes parameters, the CHANG-ES project observed 35 edge-on galaxies at two frequencies (1.6 GHz = L-band and 6 GHz = C-band) in order to probe the faint gaseous halo regions of spiral galaxies.  It was important to use more than one configuration since disk-halo structures are seen over many  spatial scales.  A summary of some selected results can be found in \cite{2019AJ....158...21I} and further information and downloadable images are at {\tt queensu.ca/changes}. Fig.~\ref{fig:median} reveals the extent and significance of gaseous halos as revealed by CHANG-ES. Since the emission is non-thermal, magnetic fields {\it must} extend out into the entire halo region shown.  It is clear that, if halos are included, spiral galaxies would look nothing like the thin flat disks that are normally depicted in standard images.

Gaseous halos are important because they provide a crucial interface between galaxy disks and the intergalactic medium (IGM).  Like the Sun that shows an abundance of activity on its surface and transitions to a Solar wind that permeates interplanetary space, so too galaxies experience much activity in their halos and also reveal winds that can exceed the escape speed, extending into the IGM \citep{2018A&A...611A..72K}.  And just like the Sun, the key and arguably most crucial ingredient is the magnetic field, its strength and topology.

CHANG-ES has made the structure of magnetic fields a priority.  Already, new results are emerging that have never before been seen.  An example is reversing rotation measures in halos \citep{Mora2019}. A model for such reversals has been developed from dynamo theory in which magnetic spiral arms are not restricted to the disk but rise into the halo regions \citep{2019MNRAS.487.1498W}. 

Galaxy halos have weak radio emission compared to disks and emission related to magnetic fields (Stokes Q and U) is weaker still.  {\it Sensitivity} is therefore mostly needed. CHANG-ES Q and U sensitivities range from about 4 to 10 $\mu$Jy/beam.  A factor of 10 in sensitivity would go a long way in answering some of the above questions.  From a technical standpoint, an easy seamless interface between single-dish (for zero-spacing flux) and interferometers would help in ensuring that all relevant spatial scales are integrated correctly into maps.  Currently, software is only being developed now to combine wide-band GBT and VLA data, and future instruments should build on these developments so that future users do not have to re-invent this wheel.

\section{\label{sec:future}Future Canadian Leadership in Magnetism Research}

Given the significant roles that Canadians do and have played in advancement in the understanding of cosmic magnetism, there is great potential to continue this leadership in the next decade and beyond. There are several upcoming projects where Canadian participation will allow us to continue in leading roles.

\subsection{Square Kilometre Array}
The Square Kilometer Array (SKA) is an international effort to build the world’s largest radio telescope, with eventually over a square kilometre of collecting area (see also WP E043). The telescope will be built in several parts, with the low frequency array being Australia and the mid-frequency array located in South Africa. The SKA has identified five key science drivers that ``aims to solve some of the biggest questions in the field of astronomy.'' One of these key science projects is \textit{The origin and evolution of cosmic magnetism} \citep[see][]{2005AAS...20713703G}. Several Canadians are members of this key SKA-Scientific Working Group. 

The SKA is anticipated to be an excellent polarization instrument, with SKA-LOW being $\sim8$ times more sensitive than LOFAR and SKA-MID $\sim5$ times more sensitive than the VLA. SKA should increase the number of RM sources by a factor of $\sim200$ from the current best available catalogue \citep{2018arXiv181003619M}.

With CIRADA, the radio community is beginning to rethink the current data processing and visualization methods and come up with new and better ways to manage extremely large data sets. The SKA Organisation has adopted a model that relies on so-called SKA Regional Centres for advanced data processing and science extraction. CIRADA will help build the Canadian capacity needed to participate in projects like the SKA Regional Centres. 

Thirteen countries are at the core of the SKA (Canada currently being one), and 100 organisations in $\sim$20 countries have been participating in the design and development of the SKA and are now engaged in the detailed design. With commissioning expected to start in the mid 2020s, there is still much work to be done in order to reach the full science of the project. Continued investment by Canada and Canadian astronomers is key to ensuring the scientific returns of this instrument, which for the field of magnetism will be unmatched by anything previous.

\subsection{CHORD: the Canadian HI Observatory and Radio transient Detector}
CHORD (see also white paper E029) will solidify Canada’s leadership in cosmic magnetism.
The long wavelengths and broad bandwidth 
of CHORD will deliver exquisite resolution in Faraday depth, yielding unprecedented views of magnetic field structures. Previous work at low frequencies has either used single dishes, with all-sky coverage but poor angular detail, or aperture synthesis, which can only focus on tiny details. CHORD will bridge the gap, revealing the role of large-scale magnetic fields in small-scale phenomena. 

CHORD will make repeated measurements (over days, weeks, or months), to enable searches for Faraday depth variability. Such variations may be inherent to the extragalactic sources, but could also be a new tool for the study of interstellar turbulence in a way that has never been done before.

With $\approx 500$ antennas packed into a $(200 \textrm{ m})^2$ area and $300-1800$~MHz frequency coverage, CHORD will complement GMIMS data (from large single antennas), and data obtained from CHIME, boosting angular resolution by factors of 3 to 5, promising breakthroughs in mapping magnetic field configuration in the Milky Way. We advocate for a future expansion of CHORD to include $\approx 80$ additional antennas spread around the DRAO site to maximum baselines of $1-2$~km and arranged to achieve good instantaneous $uv$ coverage. Such a telescope would enable an all-northern sky GMIMS survey at a resolution of a few arcminutes at the lowest frequency, enabling investigation of magnetized turbulence down to small scales.

\subsection{DRAO Synthesis telescope upgrade}

The DRAO Synthesis telescope (ST) was originally built in the 1970s to observe atomic hydrogen (at 1420 MHz). 
Later upgrades, including a correlator and spectrometer formed the basis of the Canadian Galactic Plane Survey \citep[CGPS;][]{2003AJ....125.3145T}, which ultimately revolutionized magnetic field studies by simultaneously observing polarisation angles at multiple wavelengths. This allowed for the first unambiguous determination of rotation measures for extragalatic compact sources within the Galactic disk \citep{2003ApJ...592L..29B}. 

It has been two decades since the last major upgrades to this pioneering facility.  With the support of an NSERC Collaborative Research Grant, a new correlator was been built in 2018 and is currently being tested for the ST using technology developed for CHIME (MSc thesis, P.\ Freeman). Additionally, a new multi-wavelength feed is being developed for the antennas (PhD thesis, X.\ Du). Future plans are detailed in the ST white paper (E080). The goal of these upgrades is to increase the bandwidth to cover 400-1800 MHz, making additional spectral lines observable, with broader continuum, and increased sensitivity. This will open up opportunities including RM synthesis with the interferometer, complementing the RM synthesis capabilities of the Galt Telescope. 

\subsection{DRAO John A.~Galt Telescope Upgrade}

The DRAO John A.~Galt 26-m telescope is undergoing a complete upgrade of its signal path and control system.  A MeerKAT $L$-band receiver has been purchased and fitted with ultra-low-noise cryogenically cooled amplifiers designed at the NRC.  A full-Stokes spectral line and continuum backend has been assembled using GPUs and FPGA-based CHIME IceBoards. The telescope will now be capable of producing channels of 3 Hz bandwidth across 900-1800 MHz allowing for observations of Zeeman splitting in the 21-cm hydrogen emission line, the 18-cm OH transitions, and dozens of radio recombination lines from diffuse Galactic hydrogen, helium, and carbon. Zeeman detections of 21-cm emission are notoriously difficult because instrumental polarization conversion can contribute Zeeman-like features to circular polarization spectra; this is unfortunate since 21-cm emission can allow us to probe $B$ fields in a vast volume of the Galaxy. DRAO expertise in antenna modeling \citep{Du:Landecker:2016,Robishaw:Heiles:2018} paired with the very simple geometry and optics of the Galt 26-m telescope will allow for a thorough accounting of these instrumental contributions to any detected Galactic $B$ fields.

\begin{lrptextbox}[How does the proposed initiative result in fundamental or transformational advances in our understanding of the Universe?]
%insert your text here
Magnetism contributes to every astrophysical process on every scale and its study offers considerable opportunity for transformational understanding. These include such fundamental questions as where and how did magnetic fields originate and what is their role in the evolution of the Universe. See Sec.~\ref{sec:science} for more details.

\end{lrptextbox}

\begin{lrptextbox}[What are the main scientific risks and how will they be mitigated?]
%insert your text here

Obtaining RMs of point sources (i.e., RM-grids) have proven to be extremely useful, and pose less scientific risk than diffuse emission studies. Understanding Faraday depth in extended emission and its connection to physical structures holds enormous potential as a tremendous amount of information must be encoded in the data, yet its interpretation remains a very challenging task. This can be mitigated by prioritizing experiments that have a goal to measure point sources, and for both cases, ensuring broad bandwidth to sample a wide range of Faraday scales and avoid regimes of depolarization. 

\end{lrptextbox}

\begin{lrptextbox}[Is there the expectation of and capacity for Canadian scientific, technical or strategic leadership?] 
%insert your text here
With an extremely strong record of leadership in this field, there is every expectation that Canada has the capacity to lead projects in all three areas. In terms of large scale projects, the SKA offers the greatest potential to revolutionize the study of cosmic magnetism, as well as offering a significant opportunity for Canadians to play a strategic and scientific leadership role on an international stage. Smaller scale projects serve a role in providing training opportunities for students and to maintain its leadership in engineering and technology development. See Sec.~\ref{sec:future} for further discussion of future opportunities.

\end{lrptextbox}

\begin{lrptextbox}[Is there support from, involvement from, and coordination within the relevant Canadian community and more broadly?] 
%insert your text here
The Canadian radio magnetism community is very collaborative and there are strong ties to the international community. There are also ties to and research synergies with the multi-wavelength community in Canada. 
\end{lrptextbox}

\begin{lrptextbox}[Will this program position Canadian astronomy for future opportunities and returns in 2020-2030 or beyond 2030?] 
%insert your text here
With several new telescopes currently ramping up to full capacity (e.g., ASKAP, LOFAR, MWA, MeerKAT), new projects, like CIRADA, and the SKA on the horizon, the next decade (and beyond) looks very promising for cosmic magnetism science in Canada.
\end{lrptextbox}

\begin{lrptextbox}[In what ways is the cost-benefit ratio, including existing investments and future operating costs, favourable?] 
%insert your text here
Given the international leadership potential that Canadians have demonstrated in this research area, the benefit to Canada is be expected to be high. See the SKA WP (E043) for discussion in that context. Other projects are relatively low cost compared to the opportunities for training and technology development.
\end{lrptextbox}

\begin{lrptextbox}[What are the main programmatic risks
%Instructions: Programmatic risks include but are not limited to schedule, feasibility, budget, technical readiness level, computational or software requirements, dependence on other partners, and governance plan.
and how will they be mitigated?] 
%insert your text here
The most significant programatic risks to magnetism science are availablitiy of adequate computational facilities. Both computational power and data storage are concerns as the data volumes continue to increase. Expertise in the use of these facilities is also essential to assist astronomers, whose primary role should be to investigate the science. Investment in these areas is essential to ensure continued success.

Other concerns include the technical readiness of facilities, including but not limited to their ability to accurately calibrate the data, which is particularly challenging for polarization. This can be mitigated by developing expertise to assist in assessing data quality and help provide innovative solutions. The quality of radio frequency interference environment at the facility is an additional ongoing concern.

\end{lrptextbox}

\begin{lrptextbox}[Does the proposed initiative offer specific tangible benefits to Canadians, including but not limited to interdisciplinary research, industry opportunities, HQP training,
%HQP=Highly qualified personnel, defined as individuals with university degrees at the bachelors' level and above.
EDI,
%EDI = equity, diversity and inclusion 
outreach or education?] 
%insert your text here
Not only are there many opportunities for training young astronomers in this exciting research area, there are also ample opportunities for students to gain other transferable skills. The development of new telescope technology offer opportunities for training in engineering and the extremely large data sets provide the chance to learn ``big data'' strategies and gain familiarity with high performance computing environments. 

\end{lrptextbox}

\bibliography{example} 

\end{document}